\documentclass[aps,twocolumn,showpacs,superscriptaddress,prc]{revtex4}

\usepackage{epsfig}
\usepackage{amsmath}

\begin{document}
\newcommand{\tb}{ {\bf {t}}}

\newcommand {\Co}{$^{57}$Co}

\newcommand {\Ar}{$^{39}$Ar}

\newcommand {\Na}{$^{22}$Na}

\newcommand {\Kr}{$^{83}$Kr$^{\mathrm{m}}$}

\newcommand {\mus}{$\mu$s}

\title{Calibration of liquid argon and neon detectors with \Kr}
\author{W.~H.~Lippincott}
\affiliation{Department of Physics, Yale University, New Haven, CT 06511}
\author{S.~B.~Cahn}
\affiliation{Department of Physics, Yale University, New Haven, CT 06511}
\author{D.~Gastler}
\affiliation{Department of Physics, Boston University, Boston, MA 02215}
\author{L.~W.~Kastens}
\affiliation{Department of Physics, Yale University, New Haven, CT 06511}
\author{E.~Kearns}
\affiliation{Department of Physics, Boston University, Boston, MA 02215}
\author{D.~N.~McKinsey}
\email{daniel.mckinsey@yale.edu}
\affiliation{Department of Physics, Yale University, New Haven, CT 06511}
\author{J.~A.~Nikkel}
\affiliation{Department of Physics, Yale University, New Haven, CT 06511}

\date{\today}

\begin{abstract}
We report results from tests of \Kr\, as a calibration source in liquid argon and
liquid neon. \Kr\, atoms are produced in the decay of $^{83}$Rb, and a clear
\Kr\, scintillation peak at 41.5 keV appears in both liquids when filling our
detector through a piece of zeolite coated with $^{83}$Rb. Based on this
scintillation peak, we observe 6.0
photoelectrons/keV in liquid argon with a resolution of $6\%$ ($\sigma$/E) and
3.0 photoelectrons/keV in liquid neon with a resolution of $19\%$
($\sigma$/E). The observed peak intensity subsequently
decays with the \Kr\, half-life after stopping the fill, and we find
evidence that the spatial location of \Kr\, atoms in the chamber can be
resolved. \Kr\, will be a useful calibration source for liquid argon and neon dark matter
and solar neutrino detectors. 
\end{abstract}


\maketitle

  \section{Introduction}
  \label{sec:intro}
Liquefied nobles gases are widely used as targets in low background
searches, particularly in direct dark matter searches where a weakly
interacting massive particle (WIMP) may scatter elastically with a nucleus to
produce a nuclear recoil in the liquid. Several WIMP-nucleon cross-section
limits have been set in recent years using liquid argon and xenon, and
several larger, more sensitive argon and xenon detectors are currently under
construction~\cite{Angle:2007,Alner:2005,Lebedenko:2009,Brunetti:2005}. Liquid neon is another target attracting interest, having been
proposed as a potential target for {\it pp}-solar neutrinos in addition to dark
matter~\cite{McKinsey:2000,Boulay:2006}. In particular, the MiniCLEAN detector which will be deployed at SNOLAB
in the Creighton mine in Sudbury, Ontario will run in both liquid argon and
liquid neon phases~\cite{McKinsey:2007}. 

Because the expected dark matter
signal has energies of tens of keV and drops exponentially with
increasing energy, a dark matter detector must be calibrated at
low energy to precisely determine its energy threshold and ultimate WIMP
sensitivity. Low energy calibrations are also important for a liquid neon
{\it pp} neutrino detector because any uncertainty in the energy scale produces a systematic
error in the observed neutrino energy spectrum and flux. As liquid noble gas detectors get larger, self-shielding will
render it increasingly difficult to illuminate the central volume of the liquid with
external gamma rays, particularly at low energies. Therefore, a low-energy radioactive
source that can be distributed throughout the detector volume is highly
desirable for calibrating the new generation of liquid noble gas detectors. One
example of such a calibration was the introduction of activated xenon isotopes
into the XENON10 detector. A sample of xenon was irradiated at Yale University
to produce $^{129}$Xe$^{\mathrm{m}}$ and $^{131}$Xe$^{\mathrm{m}}$, which emit 236
keV and 164 keV gamma rays with half-lives of 8.9 and 11.8 days,
respectively. This sample was shipped to Gran Sasso National Laboratory in
Italy and introduced into the detector to provide a uniform energy
calibration~\cite{Ni:2007}. While this calibration was successful, the energies of these isotopes are higher
than those expected from a WIMP signal and their relatively long half-lives
limit the frequency of deployment. 

\begin{figure}[htbp]
  \centerline{
    \hbox{\psfig{figure=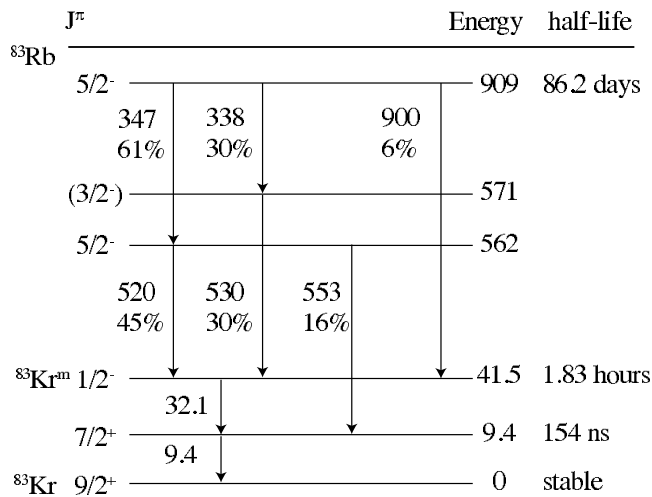,width=6.5cm,angle=270,%
                clip=}}}
          \caption[Scintillation cell]
                  {Energy levels in keV of $^{83}$Rb. $75\%$ of the time, $^{83}$Rb decays to \Kr\,,
                    which in turns emits two conversion electrons at 32.1 and
                    9.4 keV. The half-life of \Kr\, to decay via the first
                    electron to the $7/2^{-}$ state is 1.83 hours and the
                    half-life for the subsequent decay to the stable $9/2^{+}$ state of
                    $^{83}$Kr is 154 ns.}
          \label{fig:KrLevels}
\end{figure}

An alternative is the use of \Kr, which is
used as a diagnostic tool for studying the beta decay of tritium and has been
proposed for use with the KATRIN detector~\cite{Katrin:2001}. \Kr\ is produced in the decay of
$^{83}$Rb, which has a half-life of 86.2 days. In turn, the \Kr\, decays with a
half-life of $(1.83\pm0.02)$ hours, emitting 32.1 keV and 9.4 keV conversion
electrons (see Fig.~\ref{fig:KrLevels})~\cite{Venos:2005,Venos:2006,Wu:2001}. As a noble gas, \Kr\, is easily introduced into noble
liquid detectors without compromising the purity of the liquid, and it is not
expected to create any long-lived radioisotopes.  Recently, \Kr\, has been
successfully used to calibrate liquid xenon detectors~\cite{Kastens:2009,Manalaysay:2009}. Because argon
and neon are liquids at temperatures below the freezing point of krypton, any
\Kr\, atoms in the liquid could potentially freeze out on detector surfaces before decaying. This
paper describes successful tests of \Kr\, as a calibration source in both liquid
argon and liquid neon.  While some \Kr\, atoms may be freezing out,
enough reach the center of the liquid volume to provide a good energy calibration.

  \section{Experimental Apparatus}
  \label{sec:det_design}

\begin{figure}[htbp]
  \centerline{
    \hbox{\psfig{figure=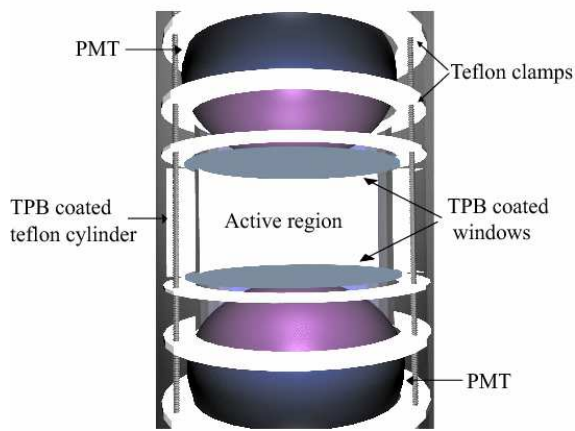,width=7.5cm,%
                clip=}}}
          \caption[Scintillation cell]
                  {(Color online) Schematic representation of the scintillation cell.}
          \label{fig:cell}
\end{figure}

A schematic of the apparatus used to perform these measurements at Yale is
shown in Fig~\ref{fig:cell}. Known as
MicroCLEAN, the apparatus has been described in detail elsewhere~\cite{Lippincott:2008}.
The detector has an active volume of 3.14 liters viewed by
two 200-mm-diameter Hamamatsu R5912-02MOD photomultiplier tubes
(PMTs), which are specifically designed for use in cryogenic
liquids. The active volume
is defined by a polytetrafluoroethylene (PTFE) cylinder 200~mm in diameter and 100~mm in height, with
two 3-mm-thick fused-silica windows at top and bottom. Both liquid
argon and neon scintillate in the ultraviolet; therefore, all inner surfaces of the
PTFE and windows are coated with $(0.20 \pm 0.01)$ mg/cm$^2$ of tetraphenyl butadiene (TPB)~\cite{McKinsey:1997},
which wavelength shifts the ultraviolet light to approximately 440~nm. The active
volume, PTFE cylinder, windows and PMTs are all contained in a stainless
steel vessel and immersed in the
liquid. The stainless steel vessel is held within a large vacuum dewar, and the system is cooled by a pulse tube refrigerator connected to a copper
liquefier. Argon or neon gas is passed through a getter for purification
before entering the vacuum dewar, passing into the liquefier and dripping into the detector.

The $^{83}$Rb source is the same source described in~\cite{Kastens:2009},
consisting of Rb-infused zeolite held in the bottom arm of a VCR cross. 700
nCi of $^{83}$Rb were loaded into the zeolite trap in February, 2009. Given
the $^{83}$Rb half-life of 86.2 days, approximately 100 nCi remained in the
trap when the tests described in this paper were performed in October and November,
2009. 

\begin{figure}[htbp]
  \centerline{
    \hbox{\psfig{figure=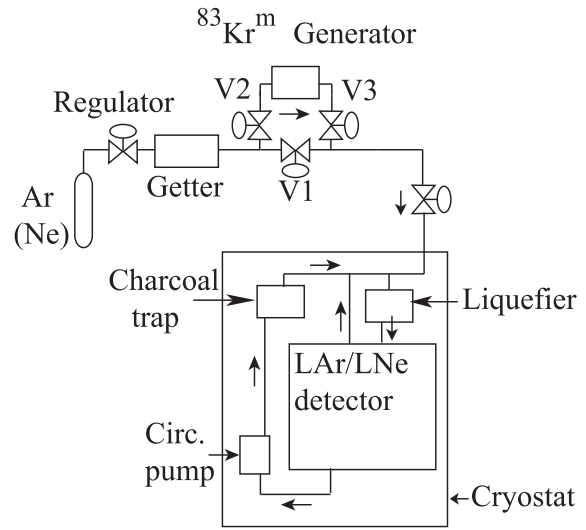,width=7.5cm,%
                clip=}}}
          \caption[Gas handling system]
                  {The gas handling system for the \Kr\, runs. In active mode, valve V1 is closed while valves
                    V2 and V3 are open. In passive mode, valves V1 and V3 are
                    open while valve V2 is closed. }
          \label{fig:gashandle}
\end{figure}

The gas handling system is shown in Fig.~\ref{fig:gashandle}. The $^{83}$Rb
trap is connected to the gas inlet line just outside the vacuum
dewar and the inlet gas can be diverted through the trap on its way into the
detector. The $^{83}$Rb remains attached to the zeolite, but the \Kr\, is free
to escape with the flowing gas into the detector. There is an additional circulation loop inside the vacuum dewar. Liquid
flows out of the bottom of the stainless steel vessel into a
nearby VCR
cross attached to a heater. The heater acts as a circulation pump by boiling
the liquid, and the resulting gas then flows up a tube through a charcoal trap before reentering the top of the
liquefier. This system was operated in two modes during the \Kr\,
tests. In all cases, the heater/circulation pump was activated to ensure mixing of the liquid
in the active region.  In
normal or ``active'' operation, argon or neon gas was flowed through the $^{83}$Rb trap
before entering the liquefier with
the bypass valve, V1, closed. A second, ``passive'' mode was intended to test whether the trap
needs to be actively in the circulation or filling path to introduce \Kr\,
into the flow. In the passive mode, valve V2 between the getter and
the trap was closed, valve V3 between the trap and the detector was open, and
the bypass valve, V1, was open.  

\subsection{Data acquisition and processing}

\begin{figure}[htbp]
  \centerline{
    \hbox{\psfig{figure=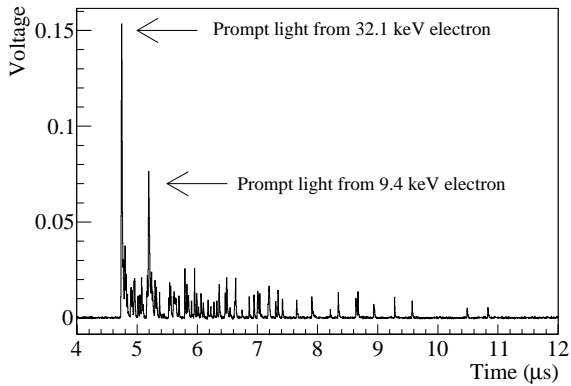,width=8cm,%
                clip=}}}
          \caption[Sample scintillation trace]
                  {Example of a \Kr\, event in liquid argon from a single PMT,
                    digitized by the 12-bit CAEN WFD, sampling at 250 MHz. The
                    32.1 and 9.4 keV components of this particular \Kr\, event are highlighted.}
          \label{fig:trace}
\end{figure}

The data acquistion system consists of a 250 MHz, 12-bit CAEN V1720 waveform
digitizer (WFD). Only the two PMT channels were recorded. Scintillation in
argon and neon is produced in the decay of metastable molecules, and there are two
decay channels with very different timing characteristics for both argon and
neon, associated with the decay of singlet and triplet molecules. For electronic recoils in argon, roughly $30\%$ of the
light comes out promptly, while the rest is distributed in time with a
lifetime of $1.5~\mu$s~\cite{Lippincott:2008}. For electronic recoils in neon,
only $10\%$ is emitted promptly, with the remainder distributed with a
lifetime of
$15~\mu$s~\cite{Nikkel:2007,Lippincott:2009}. Therefore, we collected different
record lengths of data depending on the
liquid under study; for argon, 16~\mus~ of
data were recorded for each event, while for neon, 64~\mus~were recorded. Figure~\ref{fig:trace}
shows an example \Kr\, event in argon in which the two prompt components
produced by the 32.1 and 9.4 keV electrons are highlighted. For the argon run, the
trigger rate was recorded by a counter and monitored throughout the
experiment. This counter was not available during neon running. 

Single photoelectron
spectra for the two PMTs are drawn from the tails of events as described in~\cite{Lippincott:2008}. In
neon, the gain of the PMTs drops by a factor of $\sim100$ relative to argon,
requiring the use of an additional amplifier. Also, the single photoelectron
distribution of one of the PMTs becomes too dispersed to accurately measure a
single value for the gain, and all light yield measurements are based on only one PMT. External calibration sources include 122 keV and
137 keV gamma rays from a \Co\, source, 356~keV gamma rays from a $^{133}$Ba source,
511 keV gamma rays from a $^{22}$Na source, and 662 keV gamma rays from a
$^{137}$Cs source.

The data processing, trace integration method and data cuts are similar to the
methods described
in~\cite{Lippincott:2008}. In particular, we continue to use a cut designed to eliminate events that produce light in the windows or the glass of
the PMTs by use of an asymmetry parameter, $A$. This parameter is defined as
\begin{equation}
A = \frac{S_T - S_B}{S_T + S_B},
\end{equation}
where $S_T$ and $S_B$ are the signal areas in the top and bottom PMTs, and the
value of $A$
gives a rough reconstruction of the z-position of an event in the
detector. In analysis we require that $\lvert A \rvert < 0.4$. In the argon, $95\%$ of events in the region around the observed
\Kr\, peak pass all the cuts. Because there is less prompt scintillation light in neon than
in argon and the PMT gains were greatly decreased, the trigger threshold was set extremely low for neon running,
producing a large background of noise triggers. These noise triggers are
largely eliminated by these cuts, and only $32\%$ of all events in neon in the energy
range around 41.5 keV pass the cuts.  


\section{Data Analysis and Results}
\label{sec:data_analysis}
\subsection{Liquid Argon}
\begin{figure}[htbp]
  \centerline{
    \hbox{\psfig{figure=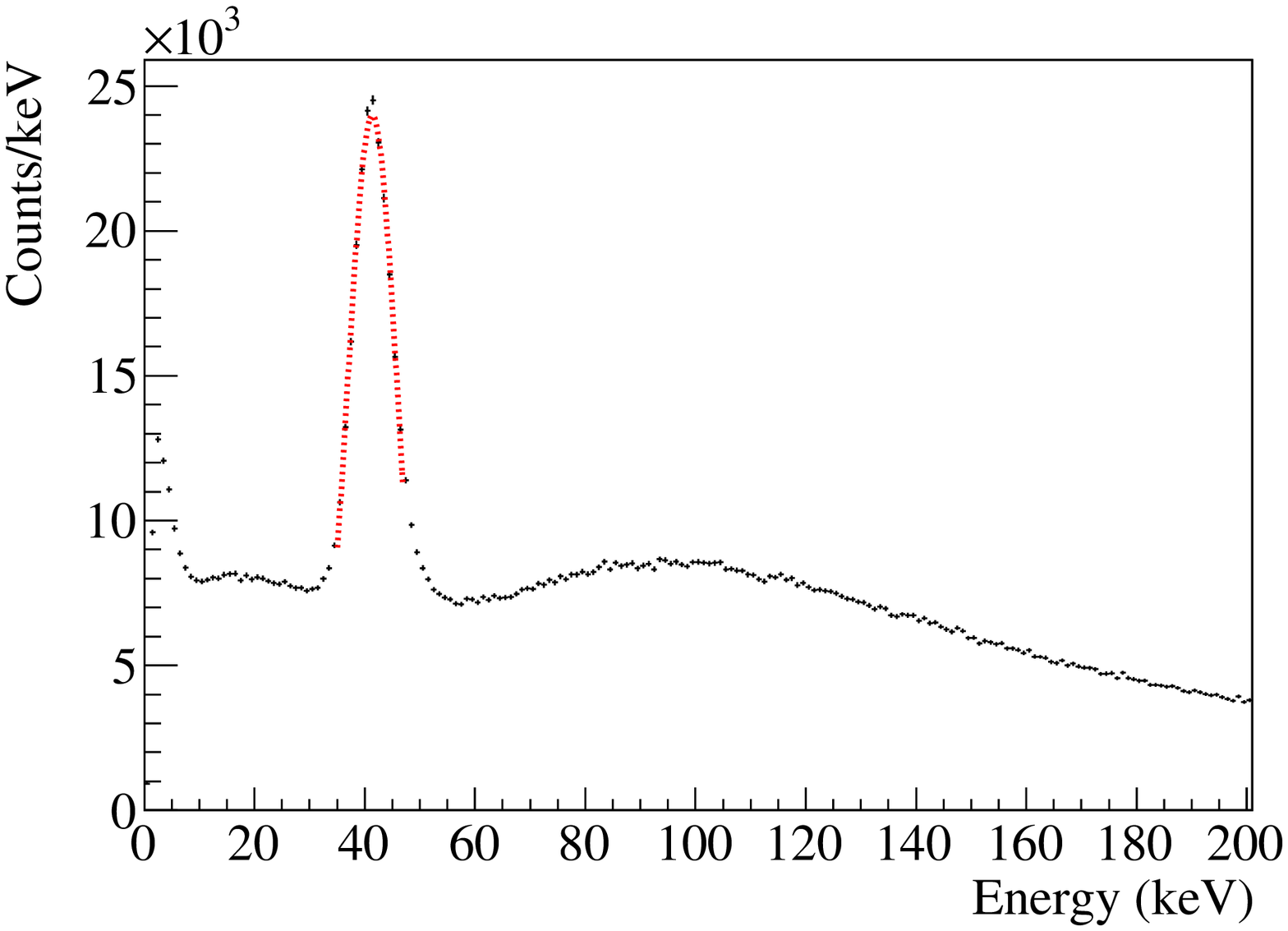,width=8cm,%
                clip=}}}
  \centerline{
    \hbox{\psfig{figure=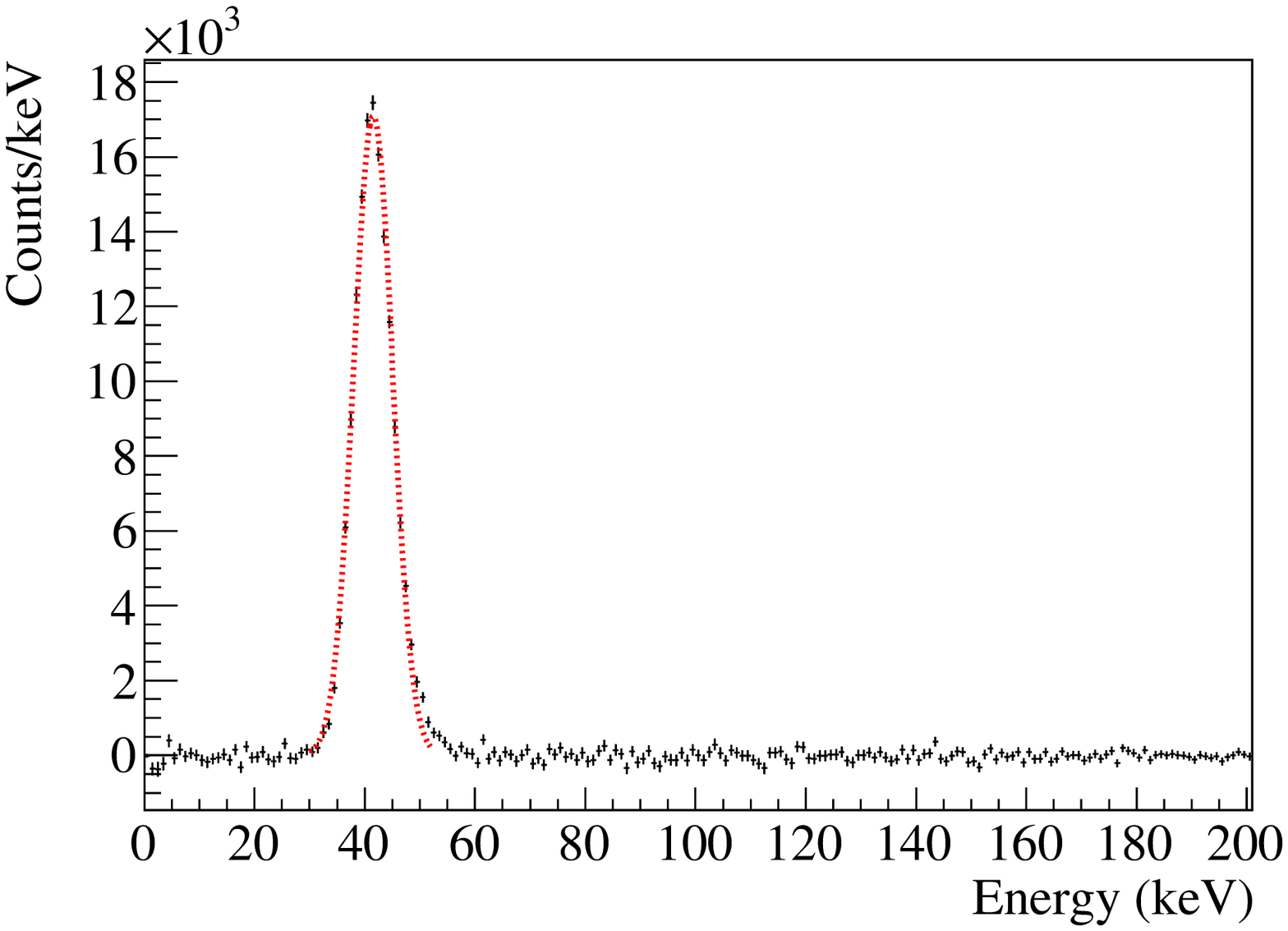, width=8cm,clip=}}}    
  \caption[Energy spectrum]
  {(Color online) Energy spectrum of \Kr\, runs in argon, with (bottom) and without
    (top) a background subtraction. The light yield is 6.0
    pe/keV and the resolution is 8.2\% ($\sigma$/E) at 41.5 keV.}
  \label{fig:argon}
\end{figure}

We performed two \Kr\, runs in argon in normal mode by filling the cell
through the \Kr\, generator for several hours before stopping the fill and
watching the decay of the introduced \Kr. We also took several background
runs to allow for a background subtraction. Running in this mode produced a clear peak in the argon, as shown in the
top panel of
Fig.~\ref{fig:argon}. After performing a background subtraction, we fit the
resulting peak to find an energy resolution of $8\%$ ($\sigma$/E) at 41.5 keV, as shown
in the lower panel of Fig.~\ref{fig:argon}. We cannot report an energy resolution at
9.4 keV; due to the timing characteristics
of scintillation in argon and neon, it is not possible to separate the light
produced by the 9.4 keV
electron from the late component of the 32.1 keV electron. 

The light yield for the detector
was $(6.0 \pm 0.2)$ photoelectrons (pe) per keV, or about 20\% higher than that observed
in~\cite{Lippincott:2008}. We attribute the increase in light yield to the use
of a
thinner layer of TPB on the walls of the PTFE cell resulting in less
absorption of the blue TPB fluorescence light by the TPB and to the substitution
of a new PMT with a slightly higher quantum efficiency. The uncertainty is
dominated by measurements of the gain in the top PMT. In terms of the total
number of photoelectrons, the energy resolution of the peak was
$1.3\times\sqrt{<N_{pe}>}$.  This analysis was repeated with a
much tighter asymmetry cut and no change was observed in the energy
resolution. As \Kr\, decays by emission of two electrons separated by a 154 ns
half-life, a second analysis was performed to try and pick out a
background-free collection of events
by looking for the characteristic double peak structure (see
Fig.~\ref{fig:trace}). Again, no improvement in energy resolution was observed
relative to the standard background subtracted analysis shown in Fig.~\ref{fig:argon}. 

\begin{figure}[htbp]
  \centerline{
    \hbox{\psfig{figure=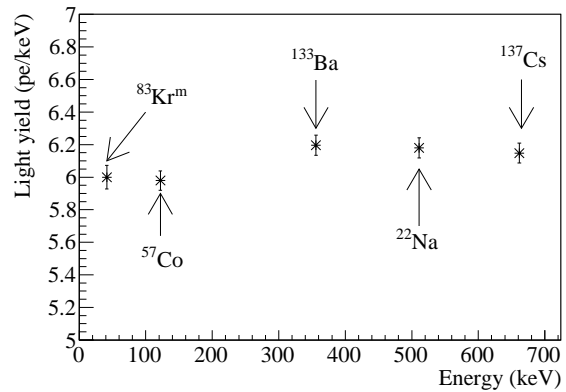,width=8cm,%
                clip=}}}    
          \caption[Energy spectrum]
                  {Light yield versus energy in argon, referenced to the value
                    of 6.0 pe/keV measured for the \Kr\, peak. There is a
                    $1\%$ systematic error on each point stemming from
                    variations in the position of the \Kr\, peak from run to
                    run.}
          \label{fig:yieldvenergy}
\end{figure}

Due to saturation of
the PMTs, the voltage must be lowered to observe high energy events; at low
voltage, the single photoelectron peak was not resolvable, although the \Kr\,
peak remained clear. We took data using the four sources
mentioned in the previous section during the steady state of a \Kr\, run. In
each case, the \Kr\, peak and the peak due to the source in use were visible
in the same data set. Given the 6.0 pe/keV observed for the
\Kr\, peak as a reference, we measured the light yield of our detector as a function
of energy to be constant to within $2\%$ between 40 and 670 keV, as shown in
Fig.~\ref{fig:yieldvenergy}.  We estimate a systematic error of $1\%$ for each point stemming from variations in the location of the \Kr\,
peak from run to run. 

\begin{figure}[htbp]
  \centerline{
    \hbox{\psfig{figure=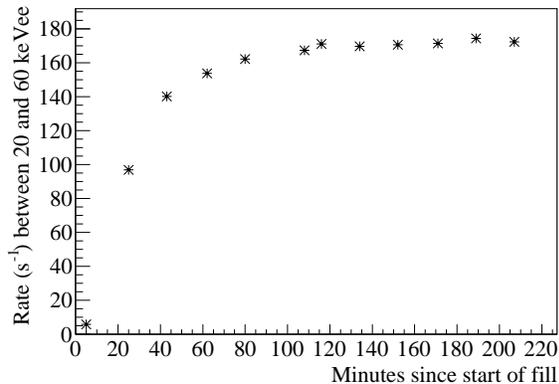,width=8cm,%
                clip=}}}    
          \caption[Energy spectrum]
                  {Rate of \Kr\, events in argon as a function of time from the
                    beginning of a fill. The rate reaches a steady state at
                    170 s$^{-1}$.}
          \label{fig:intro}
\end{figure}

\begin{figure}[htbp]
  \centerline{
    \hbox{\psfig{figure=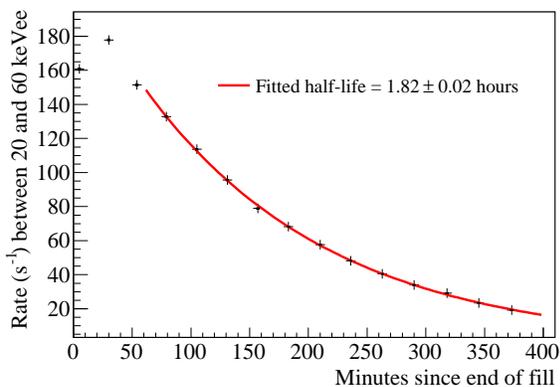,width=8cm,%
                clip=}}}    
          \caption[Energy spectrum]
                  {(Color online) Rate of \Kr\, events in argon as a function of time from ending a
                    fill. The rate decays with a fitted half-life of $(1.82 \pm 0.02)$
                    hours, consistent with the reported value of $(1.83 \pm 0.02)$ hours.}
          \label{fig:decay}
\end{figure}

Figures~\ref{fig:intro} and
~\ref{fig:decay} show the \Kr\, rate as a function of time from the start and
stop times of the fill respectively. While filling, the krypton rate slowly
rises before reaching a steady state at 171 s$^{-1}$. Given the roughly 100 nCi activity of the
Rb in the \Kr\, generator, about $6\%$ of all \Kr\, atoms created reach the active volume at
steady state in the
active mode. The active volume makes up about $10\%$ of the entire volume
of the cell, suggesting that more than half of the \Kr\, atoms produced in the trap reach the
liquid. Therefore, less than half of the \Kr\, atoms are freezing
out on the detector surfaces. As krypton binds more efficiently to charcoal
than argon, it is also likely that some fraction of the krypton is entering the charcoal trap and
freezing onto the charcoal. After stopping the
fill and allowing an hour for the detector to settle, we observed the
\Kr\, to decay with a fitted half-life of $(1.82 \pm 0.02)$ hours, consistent
with the reported half-life of $(1.83 \pm 0.02)$ hours.

\begin{figure}[htbp]
  \centerline{
    \hbox{\psfig{figure=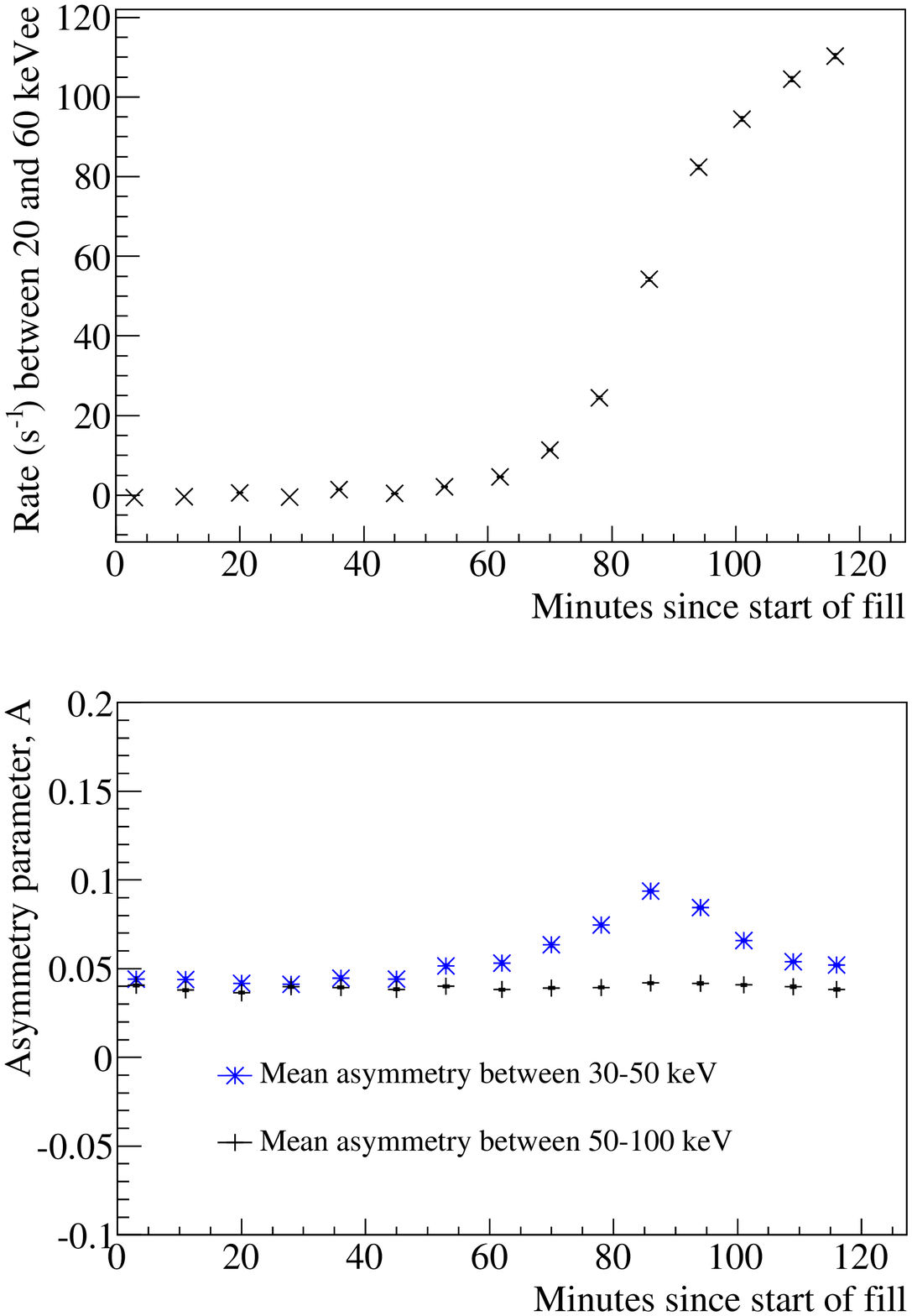,width=8cm,%
                clip=}}}    
          \caption[Energy spectrum]
                  {(Color online) The top panel shows the rate of \Kr\, events in argon as a function of time
                    from the beginning of the fill. The bottom panel
                    shows the mean value of the
                    asymmetry parameter, $A$, in argon during the same fill. The \Kr\, initially 
                    appears at the top of the detector, temporarily raising
                    $A$, before
                    filling the whole
                    active region as discussed in the text.}
          \label{fig:asymm}
\end{figure}

Figure~\ref{fig:asymm} shows the rate at the \Kr\, peak for a second run, as well as the
mean value of a Gaussian fit to the asymmetry parameter, $A$, between 30 and 50 keV and between 50 and 100 keV. The error bars from the fit are too small to
be seen on the plot. Initially, the asymmetry parameter has a slight offset
due to a relative difference in the efficiency of the PMTs. As \Kr\, begins to
enter the active region, the rate around 41.5 keV begins to increase. At the
same time, the mean value of $A$ in that region also increases, while it
remains unchanged in a different energy band. We interpret this data to
suggest that because the liquid enters the stainless steel vessel
from the top, the \Kr\, first
appears at the top of the active volume, causing an increase in the observed
asymmetry parameter that is not seen in a different energy band. As the run
continues, the \Kr\, fills out the entire active volume, and the asymmetry parameter returns to
its usual value. This illustrates the potential use of \Kr\, atoms as tracers to understand fluid flows and
mixing rates in the detector. 

\begin{figure}[htbp]
  \centerline{
    \hbox{\psfig{figure=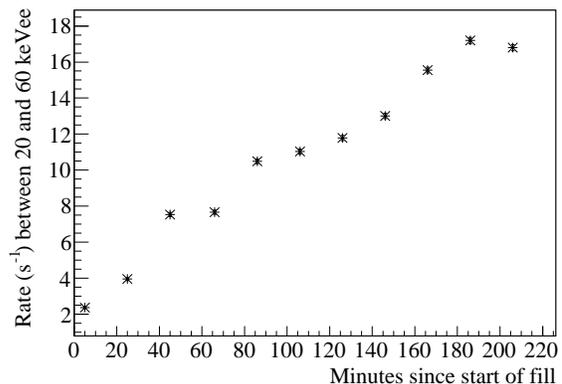,width=8cm,%
                clip=}}}    
          \caption[Energy spectrum]
                  {Rate of \Kr\, events in argon as a function of time from the
                    beginning of the fill when running in passive mode. In
                    this mode, the rate of \Kr\, is $10\%$ that observed for
                    the active mode.}
          \label{fig:intro2}
\end{figure}

We performed one run in the passive filling mode, where the gas flow did not directly
pass through the trap. Fig.~\ref{fig:intro2} shows the rate as a function of
time from beginning the fill. About one tenth the amount of
krypton entered the active region as compared to the active mode, or $0.6\%$
of all \Kr\, produced in the generator.

\subsection{Liquid neon}
\begin{figure}[htbp]
  \centerline{
    \hbox{\psfig{figure=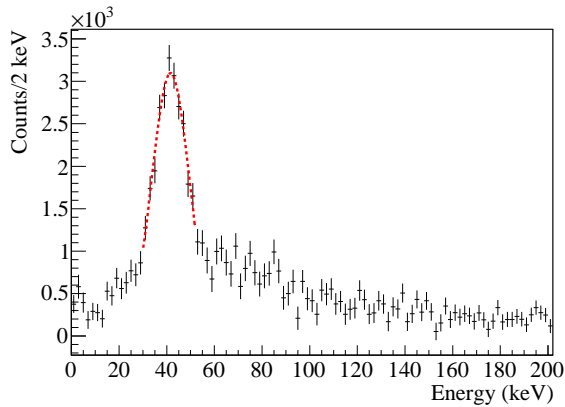, width=8cm,clip=}}}    
          \caption[Energy spectrum]
                  {(Color online) Energy spectrum of a background subtracted \Kr\ run in
                    neon. As discussed in the text, the extrapolated light
                    yield is $(3.0 \pm 0.3)$
                    pe/keV and the resolution is $19\%$ ($\sigma$/E) at 41.5 keV.}
          \label{fig:neon}
\end{figure}

We performed one run in liquid neon in active filling mode.  As mentioned in the
previous section, the trigger threshold was very low and many of the observed
events were low threshold noise events. While the background can still
be effectively subtracted, the trigger rate is dominated by these backgrounds
and the \Kr\, statistics are not nearly as good as for the argon
runs. Even so, a clear peak appears in liquid neon at the full energy of the \Kr\,
decay, as shown in Fig.~\ref{fig:neon}. We recorded a light yield of $(1.45 \pm 0.2)$
pe/keV in the bottom PMT. From the average $A$ value determined in argon from
all runs, we
determine the top tube is $6\%$ more efficient than the bottom tube. Using
the measured efficiency, we
extrapolate the total light yield in liquid neon from both tubes to be $(3.0 \pm 0.3)$
pe/keV. The error is mainly due to uncertainty in the single photoelectron
response of the bottom PMT. There is some error introduced by the
extrapolation to the second tube, as the relative efficiency of
the tubes may change between 85 K and 25 K, but this is likely to be smaller
than the error already present in the determination of the single
photoelectron response.

\begin{figure}[htbp]
  \centerline{
    \hbox{\psfig{figure=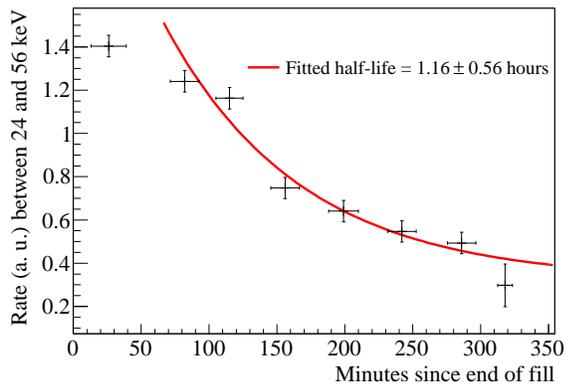,width=8cm,%
                clip=}}}    
          \caption[Energy spectrum]
                  {(Color online) Rate in arbitrary units of \Kr\, in neon as a function of time from the
                    end of a fill. The rate decays with a fitted half-life of
                    $(1.16 \pm 0.56)$ hours,
                    consistent with the value in the literature.}
          \label{fig:neondecay}
\end{figure}

 The energy resolution at 41.5 keV in
liquid neon was $19\%$ ($\sigma$/E), or $2.0\times\sqrt{<N_{pe}>}$. Fig.~\ref{fig:neondecay} shows the
decay of \Kr\, in the liquid after stopping the fill. We again wait one hour for
liquid to stop filling the detector before observing a fitted half-life of
$(1.16 \pm 0.56)$ hours. Because no counter was available during the neon run, we
 cannot estimate the efficiency with which \Kr\, atoms entered the liquid neon
volume. 


\section{Discussion}
The results discussed here show that \Kr\, is readily introduced
into both liquid argon and liquid neon volumes and could serve as a useful calibration to characterize the
scintillation signal yield of liquid argon and neon detectors at low energies. While some \Kr\, may be freezing onto
the walls, enough atoms reach the central volume to be clearly observed. A
detector with x-y position reconstruction could potentially observe \Kr\, atoms frozen to the walls as an
exterior ring of activity in the detector. The current
design for MiniCLEAN calls for a continuous purification loop, and a $^{83}$Rb
trap could be easily included in that design. Alternatively, it appears
possible to run without directly flowing gas through the trap, although this
mode is less efficient. 

In addition, the asymmetry results show that it might be possible to spatially
resolve krypton atoms as they enter the detector, providing a handle on flow,
mixing and the
spatial resolution of a large detector. 

\begin{acknowledgments}
The authors would like to thank Joseph Formaggio, who pointed them to the
papers of Venos {\it et al.} describing the preparation of $^{83}$Rb-infused
zeolite for calibration of KATRIN. This work was supported by the David and
Lucille Packard Foundation and the US Department of Energy. 
\end{acknowledgments}

\end{document}